\documentclass[prl,aps,twocolumn,showpacs,floats]{revtex4}

\usepackage{graphicx}
\usepackage{dcolumn}
\usepackage{bm}
\usepackage{epstopdf}

\newcommand{\bq}{\begin{equation}}
\newcommand{\eq}{\end{equation}}
\newcommand{\bqa}{\begin{eqnarray}}
\newcommand{\eqa}{\end{eqnarray}}
\newcommand{\nn}{\nonumber \\}

\def\be     {\begin{equation}}
\def\ee     {\end{equation}}
\def\bea        {\begin{eqnarray}}
\def\eea        {\end{eqnarray}}
\def\bnn    {\begin{eqnarray*}}
\def\enn    {\end{eqnarray*}}

\begin{document}

\title{Novel duality in disorder driven local quantum criticality}

\author{Minh-Tien Tran$^{1,2}$ and Ki-Seok Kim$^{1,3}$}
\affiliation{
$^1$Asia Pacific Center for Theoretical Physics, POSTECH, Pohang, Gyeongbuk 790-784, Republic of Korea \\
$^2$Institute of Physics, Vietnamese Academy of Science and
Technology, P.O.Box 429, 10000 Hanoi, Vietnam \\ $^3$Department of
Physics, POSTECH, Pohang, Gyeongbuk 790-784, Korea }

\begin{abstract}
We find that competition between random Kondo and random magnetic
correlations results in a quantum phase transition from a local
Fermi liquid to a spin liquid. The local charge susceptibility
turns out to have exactly the same critical exponent as the local
spin susceptibility, suggesting novel duality between the Kondo
singlet phase and the critical local moment state beyond the
Landau-Ginzburg-Wilson symmetry breaking framework. This leads us
to propose an enhanced symmetry at the local quantum critical
point, described by an O(4) vector for spin and charge. The
symmetry enhancement serves mechanism of electron
fractionalization in critical impurity dynamics, where such
fractionalized excitations are identified with topological
excitations.
\end{abstract}


\maketitle

To find mechanism of quantum number fractionalization has been an
important direction of research in condensed matter physics
\cite{Fractionalization}. An interesting proposal is spin
fractionalization in the vicinity of a generic second order
transition between symmetry unrelated orders, forbidden in the
Landau-Ginzburg-Wilson (LGW) symmetry breaking framework
\cite{Senthil_DQCP}. An essential ingredient is that the original
symmetry of a microscopic model is enhanced at the quantum
critical point (QCP) \cite{Tanaka_SO5}, allowing the LGW forbidden
duality between the symmetry unrelated orders \cite{Momoi}. The
underlying mechanism is emergence of a topological term
\cite{Senthil_DQCP}, originating from the enhanced symmetry at the
QCP \cite{Tanaka_SO5}. The topological phase assigns a non-trivial
quantum number for the other phase to a topological soliton, the
disorder parameter of one phase, and condensation of topological
excitations results in the other phase, that is, emergence of
duality. Such topological excitations are identified with
fractionalized excitations at the deconfined QCP.

The generic second order transition between symmetry unrelated
orders was also proposed in the context of heavy fermion quantum
criticality, where a continuous transition seems to appear from an
antiferromagnetic phase to a heavy fermion Fermi liquid
\cite{QCP_Review}. Although there is no concrete construction for
an effective field theory \cite{KSKim}, an extended dynamical
mean-field theory (DMFT) shows that the local spin susceptibility
diverges at the same time as the antiferromagnetic spin
susceptibility \cite{EDMFT}, implying an enhanced emergent
symmetry because Kondo fluctuations are equally critical with
antiferromagnetic ones.

In this letter we investigate the role of strong randomness in
both Kondo and Ruderman-Kittel-Kasuya-Yosida (RKKY) correlations,
where strong disorder leads us to the DMFT approximation naturally
because disorder average covers spatial correlations
\cite{SG_Review}. We find that competition between random Kondo
and random RKKY interactions results in a quantum phase transition
from a local Fermi liquid to a spin liquid. The local charge
susceptibility turns out to have exactly the same critical
exponent as the local spin susceptibility, suggesting novel
duality between the Kondo singlet phase and the critical local
moment state beyond the LGW framework. This leads us to propose an
enhanced symmetry at the local QCP, described by an O(4) vector
for spin and charge. We argue that the symmetry enhancement serves
mechanism of impurity fractionalization at the local QCP, where
spinons are identified with instantons in an O(4) nonlinear
$\sigma$ model type on a nontrivial manifold.


The previous works focused on the origin of non-Fermi liquid
physics in the Kondo singlet phase without RKKY correlations
\cite{Kondo_Disorder}. The role of non-random RKKY interactions
was examined in Ref. \cite{Kondo_RKKY_Disorder}, where proximity
of the Anderson localization for conduction electrons was proposed
to be mechanism of the local Fermi liquid to spin liquid
transition.
We also point out the previous work on the disordered t-J model,
where holes are doped into the spin liquid state of the disordered
Heisenberg model, giving rise to marginal Fermi liquid
phenomenology \cite{Olivier}.

We start from an effective Anderson lattice model \bqa && H =
\sum_{ij,\sigma} t_{ij} c^{\dagger}_{i\sigma} c_{j\sigma} + E_{d}
\sum_{i\sigma} d^{\dagger}_{i\sigma} d_{i\sigma} + \sum_{ij}
J_{ij} \mathbf{S}_{i} \cdot \mathbf{S}_{j}  \nonumber \\
&& + \sum_{i\sigma} (V_{i} c^{\dagger}_{i\sigma} d_{i\sigma} +
{\rm H.c.}) , \eqa where $t_{ij} = - \frac{t}{M \sqrt{z}}$ is a
hopping integral for conduction electrons and \bqa && J_{ij} =
\frac{J}{\sqrt{z M}} \varepsilon_{i}\varepsilon_{j} , ~~~~~ V_{i}
= \frac{V}{\sqrt{M}} \varepsilon_{i} \nonumber \eqa are random
RKKY and hybridization coupling constants, respectively. Here, $M$
is the spin degeneracy. Randomness is given by \bqa
\overline{\varepsilon_{i}} = 0 , ~~~~~
\overline{\varepsilon_{i}\varepsilon_{j}} = \delta_{ij} . \eqa

This model has two well known limits. In the $V \rightarrow 0$
limit the random Heisenberg model results, where a spin glass
phase turns out to be unstable against a spin liquid state due to
quantum fluctuations \cite{Sachdev_SG}. In the $J \rightarrow 0$
limit the disordered Anderson lattice model was intensively
investigated, as mentioned in the introduction, where the role of
randomness in the energy level for localized electrons was
revealed \cite{Kondo_Disorder}. It is natural to expect a quantum
phase transition from the Kondo singlet phase to the spin liquid
state, increasing the ratio $V/J$.

The disorder average can be performed in the replica trick
\cite{SG_Review}. We observe that such an average neutralizes
spatial correlations except the hopping term of conduction
electrons. This leads us to the DMFT formulation
\cite{Supplementary}. Performing the DMFT approximation with the
disorder average in the replica trick, we reach an effective local
action for the strong random Anderson lattice model \bqa &&
S^{replica}_{dmft} = \int_{0}^{\beta} d\tau \Bigl\{ \sum_{\sigma
a} c^{\dagger a}_{\sigma}(\tau) (\partial_{\tau} - \mu)
c^{a}_{\sigma}(\tau) \nn && + \sum_{\sigma a}d^{\dagger
a}_{\sigma}(\tau) (\partial_{\tau} + E_d) d^{a}_{\sigma}(\tau)
\Bigr\} \nn && -\frac{V^2}{2 M} \int_{0}^{\beta} d\tau
\int_{0}^{\beta} d\tau' \sum_{\sigma \sigma' a b} \big[ c^{\dagger
a}_{\sigma}(\tau) d^{a}_{\sigma}(\tau) + d^{\dagger
a}_{\sigma}(\tau) c^{a}_{\sigma}(\tau)\big] \nn && \big[
c^{\dagger b}_{\sigma'}(\tau') d^{b}_{\sigma'}(\tau') + d^{\dagger
b}_{\sigma'}(\tau') c^{b}_{\sigma'}(\tau')\big] \nn && -
\frac{J^2}{2 M} \int_{0}^{\beta} d\tau \int_{0}^{\beta} d\tau'
\sum_{ab} \sum_{\alpha\beta\gamma\delta} S^{a}_{\alpha\beta}(\tau)
R^{ab}_{\beta\alpha\gamma\delta}(\tau-\tau')
S^{b}_{\delta\gamma}(\tau') \nn && + \frac{t^2}{M^2}
\int_{0}^{\beta} d\tau \int_{0}^{\beta} d\tau' \sum_{ab\sigma}
c^{\dagger a}_{\sigma}(\tau) G^{ab}_{c \;
\sigma\sigma}(\tau-\tau') c^{b}_{\sigma}(\tau' ) ,  \eqa where
$R^{ab}_{\beta\alpha\gamma\delta}(\tau-\tau')$ is the local
spin-spin correlation function and $G^{ab}_{c \;
\sigma\sigma}(\tau-\tau')$ is the local electron propagator.
Self-consistency is imposed in the Bethe lattice with an infinite
number of lattice coordinations \cite{DMFT_Review}.


We solve the effective local action based on the U(1) slave-boson
representation \bqa && d^{a}_{\sigma} = \hat{b}^{\dagger a}
f^{a}_{\sigma} , ~~~~~ S_{\sigma\sigma'}^{a} =
f^{a\dagger}_{\sigma} f_{\sigma'}^{a} - q_{0}^{a} \delta_{\sigma
\sigma'}  \eqa with the single occupancy constraint
$b^{a\dagger}(\tau)b^{a}(\tau) + \sum_{\sigma}
f^{a\dagger}_{\sigma}(\tau) f^{a}_{\sigma}(\tau) = 1$, where
$q_{0}^{a} = \sum_{\sigma}f^{a\dagger}_{\sigma} f_{\sigma}^{a}/M
$. First, we show existence of a quantum phase transition based on
the mean-field approximation, valid deep inside each stable phase.
Second, we reveal the nature of the QCP beyond the mean-field
approximation, where quantum corrections are introduced fully
self-consistently, justified in the $M \rightarrow \infty$ limit.
In this study we are allowed to consider only paramagnetic and
replica symmetric phases, protected due to strong quantum
fluctuations of spin $1/2$, consistent with the previous studies
\cite{Sachdev_SG}.


In the slave-boson mean-field approximation we replace the holon
operator with its expectation value $\langle \hat{b}^{a} \rangle
\equiv b^{a}$. Then, we reach self-consistent equations for
self-energy corrections in the replica symmetric phase \bqa &&
\Sigma_{c}(i\omega_l) = \frac{V^2}{M} G_{f}(i\omega_l) |b|^2 +
\frac{t^2}{M^2} G_{c}(i\omega_l) , \nn && \Sigma_{f}(i\omega_l) =
\frac{V^2}{M} G_{c}(i\omega_l) |b|^2 + \frac{J^2}{2 M} T \sum_{s}
\sum_{\nu_m} G_{f}(i\omega_l-i\nu_m) \nn && [R_{s\sigma\sigma
s}(i\nu_m) + R_{\sigma s s\sigma}(-i\nu_m) ] , \nn &&
\Sigma_{cf}(i\omega_l) = \frac{V^2}{M} G_{fc}(i\omega_l) (b^2)^{*}
- n \frac{V^2}{M} (b^2)^{*} \sum_s  \langle f^{\dagger}_{s} c_{s}
+ c^{\dagger}_{s} f_{s} \rangle , \nn && \Sigma_{fc}(i\omega_l) =
\frac{V^2}{M} G_{cf}(i\omega_l) b^2 - n \frac{V^2}{M} b^2 \sum_s
\langle f^{\dagger}_{s} c_{s} + c^{\dagger}_{s} f_{s} \rangle ,
\nn && R_{\sigma s s \sigma}(i\nu_{m}) = - \frac{1}{\beta}
\sum_{\omega_{l}} G_{f \sigma}(i\nu_{m} + i\omega_{l}) G_{f
s}(i\omega_{l}) , \eqa where the Green's functions are given by
\bqa &&
\left(\begin{array}{cc} G_{c}(i \omega_l) & G_{fc}(i \omega_l) \\
G_{cf}(i \omega_l) & G_{f}(i \omega_l)
\end{array} \right) \nn && = \left( \begin{array}{cc}
i\omega_l + \mu - \Sigma_{c}(i \omega_l) & - \Sigma_{cf}(i
\omega_l) \\
- \Sigma_{fc}(i \omega_l) & i\omega_l - E_d -\lambda -
\Sigma_{f}(i \omega_l)
\end{array} \right)^{-1} . \nonumber
\eqa $n$ is the replica index, set to be zero. Self-consistent
equations for holon condensation and an effective chemical
potential $\lambda$ are \bqa && b \Big[ \lambda + 2 V^2 T
\sum_{\omega_l} G_{c}(i\omega_l) G_{f}(i\omega_l) \nn && + V^2 T
\sum_{\omega_l} \Bigl\{ G_{fc}(i\omega_l) G_{fc}(i\omega_l) +
G_{cf}(i\omega_l) G_{cf}(i\omega_l)\Bigr\} \Big] =0 , \nn && |b|^2
+ \sum_{\sigma} \langle f^{\dagger}_{\sigma} f_{\sigma} \rangle =
1 . \eqa

The main difference between the clean and disordered cases is that
the off diagonal Green's function $G_{fc}(i\omega_l)$ should
vanish in the presence of randomness in $V$ with its zero mean
value while it is proportional to the condensation $b$ when the
average value of $V$ is finite \cite{KLM}. In the present
situation we find $b^{a} = \langle f^{a\dagger}_{\sigma}
c_{\sigma}^{a} \rangle = 0$ while $(b^{a})^{*}b^{b} = \langle
f^{a\dagger}_{\sigma} c_{\sigma}^{a} c_{\sigma'}^{b\dagger}
f_{\sigma'}^{b} \rangle \equiv |b|^{2} \delta_{ab} \not= 0$. This
implies that the Kondo singlet phase is not characterized by the
holon condensation but described by finite density of holons. It
is important to notice that this gauge invariant order parameter
does not cause any kinds of symmetry breaking for the Kondo
effect. This cures the artificial finite temperature transition in
the slave-boson mean-field theory of the Anderson lattice model
without randomness \cite{KLM}.

\begin{figure}[h]
\includegraphics[width=0.48\textwidth]{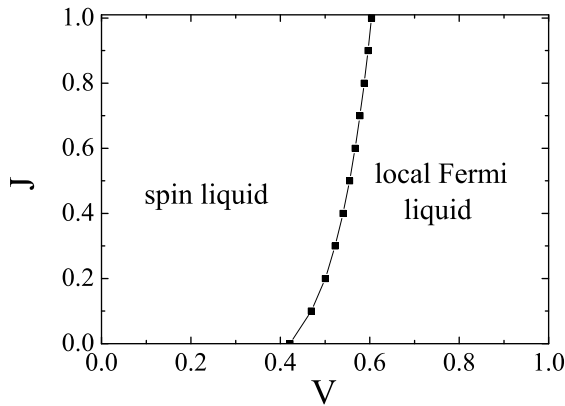}
\caption{The phase diagram of the strongly disordered Anderson
lattice model in the DMFT approximation ($E_d=-1$, $\mu=0$,
$T=0.01$, $t=1$, $M=2$).} \label{fig1}
\end{figure}

\begin{figure}[h]
\includegraphics[width=0.48\textwidth]{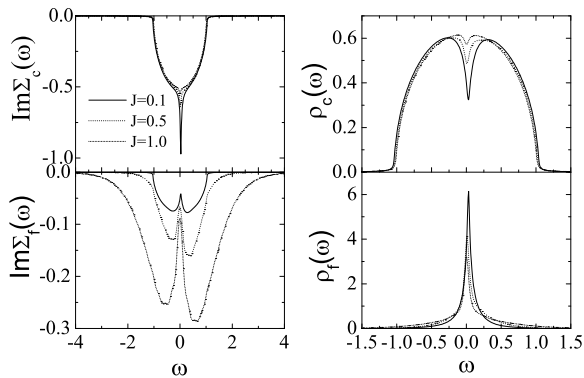}
\caption{Left: The imaginary part of the self-energy of conduction
electrons and that of localized electrons for various values of
$J$. Right: Density of states of conduction ($\rho_{c}(\omega)$)
and localized ($\rho_{f}(\omega)$) electrons for various values of
$J$. We used $V=0.5$, $E_d=-0.7$, $\mu=0$, $T=0.01$, $t=1$, and
$M=2$. } \label{fig2}
\end{figure}

Figure \ref{fig1} shows the phase diagram in the plane of $(V,
J)$, where $V$ and $J$ are variances for the Kondo and RKKY
interactions, respectively. The phase boundary is characterized by
$|b|^{2} = 0$, below which $|b|^{2} \not= 0$ appears to cause
effective hybridization between conduction electrons and localized
fermions. In the left panel of Fig. \ref{fig2} one finds that the
effective hybridization enhances the scattering rate of conduction
electrons dramatically around the Fermi energy while the
scattering rate for localized electrons becomes reduced at the
resonance energy.
%
%
%
This self-energy effect reflects the spectral function in the
right panel of Fig. \ref{fig2}, where the pseudogap feature arises
in conduction electrons while the sharply defined peak appears in
localized electrons, identified with the Kondo resonance although
the description of the Kondo effect differs from the clean case.
Increasing the RKKY coupling, the Kondo effect is suppressed as
expected. In the Kondo singlet phase the local spin susceptibility
shows the typical $\omega$-linear behavior in the low frequency
limit, nothing but the Fermi liquid physics for spin correlations.
Increasing $J$, incoherent spin correlations are enhanced,
consistent with spin liquid physics.


The nature of the local QCP is uncovered in the non-crossing
approximation, exact in the $M \rightarrow \infty$ limit
\cite{Hewson_Book}. Self-consistent equations for self-energy
corrections are
\begin{eqnarray}
\Sigma_{c}(\tau) &=& \frac{V^2}{M} G_{f}(\tau) G_{b}(-\tau) +
\frac{t^2}{M^2} G_{c}(\tau) , \nn \Sigma_{f}(\tau) &=&
\frac{V^2}{M} G_{c}(\tau) G_{b}(\tau) - J^2 [G_{f}(\tau)]^2
G_{f}(-\tau) , \nn \Sigma_{b}(\tau) &=& V^2 G_{c}(-\tau)
G_{f}(\tau) ,
\end{eqnarray} where the holon propagator is $G_{b}(i \nu_{l}) = \Big(
i\nu_{l} -\lambda -\Sigma_{b}(i\nu_l) \Big)^{-1}$.
%
%
When the second terms are neglected in the first and second
equations, these are reduced to those of the multi-channel Kondo
effect \cite{Hewson_Book}. Power-law solutions are well known in
the regime of $1/T_K \ll \tau \ll \beta=1/T$, where $T_{K} =
D[\Gamma_{c}/\pi D]^{1/M} \exp[\pi E_{d}/M \Gamma_{c}]$ is an
effective Kondo temperature \cite{Tien_Kim} with the conduction
bandwidth $D$ and effective hybridization $\Gamma_{c} = \pi
\rho_{c} \frac{V^{2}}{M}$. The RKKY interaction will reduce the
effective hybridization, where $\Gamma_{c}$ is replaced with
$\Gamma_{c}^{J} \approx \pi \rho_{c} (\frac{V^{2}}{M} - J^{2})$.

Our power-law ansatz is as follows
\begin{eqnarray}
&& G_{c}(\tau) = A_{c} \beta^{-\Delta_{c}}
g_{c}\Big(\frac{\tau}{\beta} \Big) , ~~~ G_{f}(\tau) = A_{f}
\beta^{-\Delta_{f}} g_{f}\Big(\frac{\tau}{\beta} \Big) , \nn &&
G_{b}(\tau) = A_{b} \beta^{-\Delta_{b}}
g_{b}\Big(\frac{\tau}{\beta} \Big) ,
\end{eqnarray}
where $g_{\alpha}(x) = \bigg(\frac{\pi}{\sin(\pi
x)}\bigg)^{\Delta_\alpha}$ is the scaling function \cite{Tien_Kim}
at finite temperatures with $\alpha=c,f,b$. $A_{c}$, $A_{f}$, and
$A_{b}$ are numerical constants. Inserting these expressions into
Eq. (7), we find two fixed point solutions. One coincides with the
multi-channel Kondo effect, given by $\Delta_{c} = 1$, and
$\Delta_{f} = \frac{M}{M+1}$, $\Delta_{b} = \frac{1}{M+1}$ with $M
= 2$, where contributions from spin fluctuations to self-energy
corrections are irrelevant, compared with holon fluctuations. The
other is $\Delta_{c} = 1$ and $\Delta_{f} = \Delta_{b} =
\frac{1}{2}$, where spin correlations are critical as much as
holon fluctuations. One can understand the critical exponent
$\Delta_{f} = 1/2$ as the proximity of the spin liquid physics
\cite{Sachdev_SG}. Actually, we find the local spin susceptibility
$\Im \chi(\omega) = A_{f}^{2} \Bigl( \frac{1}{T}
\Bigr)^{1-2\Delta_{f}} \Phi\Bigl(\frac{\omega}{T}\Bigr)$, where
the scaling function is \bqa \Phi(x) = 2 (2\pi)^{2\Delta_{f}-1}
\sinh\Bigl(\frac{x}{2}\Bigr)
\frac{\Gamma\Bigl(\Delta_{f}+i\frac{x}{2\pi}\Bigr)
\Gamma\Bigl(\Delta_{f}-i\frac{x}{2\pi}\Bigr)}{\Gamma
(2\Delta_{f})} , \nonumber \eqa which coincides with the spin
spectrum of the spin liquid state when $V = 0$. In this respect
the second fixed point is the genuine critical solution for the
local Fermi liquid to spin liquid transition.

It is straightforward to see that the critical exponent of the
local spin susceptibility is exactly the same as that of the local
charge susceptibility ($2\Delta_{f} = 2\Delta_{b} = 1$),
proportional to $1/\tau$.
%
%
Such an unexpected scaling behavior proposes an enhanced symmetry,
allowing us to construct an effective local field theory in terms
of an O(4) vector $\boldsymbol{\Psi}^{a}(\tau) = \left(
\begin{array}{c} \boldsymbol{S}^{a}(\tau) \\ \rho^{a}(\tau)
\end{array} \right)$,
\bqa Z_{eff} &=& \int D \boldsymbol{\Psi}^{a}(\tau)
\delta\Bigl(|\boldsymbol{\Psi}^{a}(\tau)|^{2} - 1\Bigr) e^{-
\mathcal{S}_{eff}} , \nn \mathcal{S}_{eff} &=& -
\frac{g^{2}}{2M}\int_{0}^{\beta} d \tau \int_{0}^{\beta} d \tau'
\boldsymbol{\Psi}^{a T}(\tau)
\boldsymbol{\Upsilon}^{ab}(\tau-\tau')
\boldsymbol{\Psi}^{b}(\tau') \nn &+& \mathcal{S}_{top} ,  \eqa
where $\boldsymbol{\Upsilon}^{ab}(\tau-\tau')$ determines dynamics
of the O(4) vector, resulting from spinon and holon dynamics in
principle. $g \propto V/J$ is an effective coupling constant, and
$\mathcal{S}_{top}$ is a possible topological term.

One can represent the O(4) vector generally as follows \bqa
\boldsymbol{\Psi}^{a} : \tau \longrightarrow \Bigl( && \sin
\theta^{a}(\tau) \sin \phi^{a}(\tau) \cos \varphi^{a}(\tau) , \nn
&& \sin \theta^{a}(\tau) \sin \phi^{a}(\tau) \sin
\varphi^{a}(\tau) , \nn && \sin \theta^{a}(\tau) \cos
\phi^{a}(\tau) , \cos \theta^{a}(\tau) ~~ \Bigr) ,  \nonumber \eqa
where $\theta^{a}(\tau), \phi^{a}(\tau), \varphi^{a}(\tau)$ are
three angle coordinates for the O(4) vector. It is essential to
observe that the target manifold for the O(4) vector is not a
simple sphere type, but more complicated because the last
component of the O(4) vector is the charge density field. Its
positiveness results in a periodicity, given by
$\boldsymbol{\Psi}^{a}(\theta^{a},\phi^{a},\varphi^{a}) =
\boldsymbol{\Psi}^{a}(\pi - \theta^{a},\phi^{a},\varphi^{a})$.
This folded space structure allows a nontrivial topological
excitation. Suppose the boundary configuration of
$\boldsymbol{\Psi}^{a}(0,\phi^{a},\varphi^{a}; \tau = 0)$ and
$\boldsymbol{\Psi}^{a}(\pi,\phi^{a},\varphi^{a}; \tau = \beta)$,
connected by $\boldsymbol{\Psi}^{a}(\pi/2,\phi^{a},\varphi^{a}; 0
< \tau < \beta)$. Interestingly, this configuration is {\it
topologically} distinguishable from the configuration of
$\boldsymbol{\Psi}^{a}(0,\phi^{a},\varphi^{a}; \tau = 0)$ and
$\boldsymbol{\Psi}^{a}(0,\phi^{a},\varphi^{a}; \tau = \beta)$ with
$\boldsymbol{\Psi}^{a}(\pi/2,\phi^{a},\varphi^{a}; 0 < \tau <
\beta)$ because of the folded structure. This topological
excitation carries a spin quantum number $1/2$ in its core, given
by $\boldsymbol{\Psi}^{a}(\pi/2,\phi^{a},\varphi^{a}; 0 < \tau <
\beta) = \Bigl( \sin \phi^{a}(\tau) \cos \varphi^{a}(\tau) , \sin
\phi^{a}(\tau) \sin \varphi^{a}(\tau) , \cos \phi^{a}(\tau) , 0
\Bigr)$. This is the spinon excitation, described by an O(3)
nonlinear $\sigma$ model with the nontrivial spin correlation
function $\boldsymbol{\Upsilon}^{ab}(\tau-\tau')$, where the
topological term is reduced to the single spin Berry phase term in
the instanton core \cite{Tanaka_SO5}.

In this local impurity picture the local Fermi liquid phase is
described by gapping of instantons while the spin liquid state is
characterized by condensation of instantons. Of course, the low
dimensionality does not allow condensation, resulting in critical
dynamics for spinons. This scenario clarifies the LGW forbidden
duality between the Kondo singlet and the critical local moment
for the impurity state, allowed by the presence of the
topological term.

We explicitly checked that the similar result can be found in the
extended DMFT for the clean Kondo lattice model, where two fixed
point solutions are allowed \cite{EDMFT_NCA}. One is the same as
the multi-channel Kondo effect and the other is essentially the
same as the second solution in this paper. In this respect we
believe that the present scenario works in the extended DMFT
framework although applicable to only two spatial dimensions
\cite{EDMFT}.

One may suspect the applicability of the DMFT framework for this
disorder problem. However, the hybridization term turns out to be
exactly local in the case of strong randomness while the RKKY term
is safely approximated to be local for the spin liquid state,
expected to be stable against the spin glass phase in the quantum
spin case \cite{Supplementary}. This situation should be
distinguished from the clean case, where the DMFT approximation
causes several problems such as the stability of the spin liquid
state and strong dependence of the dimension of spin dynamics
\cite{Olivier,EDMFT,EDMFT_NCA}.

In conclusion, we proposed novel duality between the Kondo and
critical local moment phases in the strongly disordered Anderson
lattice model. This duality serves mechanism of impurity
fractionalization at the local QCP, where spinons are identified
with instantons in an O(4) nonlinear $\sigma$ model on a
nontrivial manifold.



This work was supported by the National Research Foundation of
Korea (NRF) grant funded by the Korea government (MEST) (No.
2010-0074542). One of the authors (M.-T.) was also supported by
the Vietnamese NAFOSTED.

\end{document}